# Sub-Hertz resonance by weak measurement

Weizhi Qu[1], Jian Sun[1], Shenchao Jin[1], Liang Jiang[2], Jianming Wen[3*], and Yanhong Xiao[1,4*]

[1]Department of Physics, State Key Laboratory of Surface Physics, and Key Laboratory of Micro and Nano Photonic Structures (Ministry of Education), Fudan University, Shanghai 200433, China
[2]Department of Applied Physics, Yale University, New Haven, Connecticut 06511, USA
[3]Department of Physics, Kennesaw state University, Marietta, Georgia 30060, USA
[4]Collaborative Innovation Center of Advanced Microstructures, Nanjing 210093, China
*emails: jianming.wen@kennesaw.edu; yxiao@fudan.edu.cn.

**Abstract:** Weak measurement (WM) with state pre- and post-selection can amplify otherwise undetectable small signals and thus promise great potentials in precision measurements. Although frequency measurements offer the hitherto highest precision owing to stable narrow atomic transitions, it remains a long-standing interest to develop new schemes to further escalate their performance. Here, we propose and demonstrate a WM-enhanced spectroscopy technique which is capable of narrowing the resonance to 0.1 Hz in a room-temperature atomic vapor cell. Potential of this technique for precision measurement is demonstrated through weak magnetic field sensing. By judiciously pre- and post-selecting frequency-modulated input and output optical states in a nearly-orthogonal manner, a sensitivity of 7 fT/$\sqrt{\text{Hz}}$ near DC is achieved, using only one laser beam of 15 μW power. Additionally, our results extend the WM framework to a non-Hermitian Hamiltonian, and shed new light in metrology and bio-magnetic field sensing applications.

Measurement is the basis for the practice of science. Being a hallmark of quantum mechanics, this assumes even a more fundamental role since the very act of measuring a system is irrevocably accompanied by a complementary disturbance. As prototypically modeled by von Neumann, the standard measurement process, in which a quantum "system" of interest is strongly coupled to an external measuring device (or "pointer") with small uncertainty, irreversibly collapses the system into an eigenstate of the Hamiltonian operator associated with the observable, and yields its corresponding eigenvalue. Contrary to this strong (projective) procedure, the notion of weak measurements (WMs) introduced by Aharonov, Albert, and Vaidman [1] describes an intriguing situation where partial information is gained by feebly probing the system without undermining its initial state. Although the uncertainty in each measurement is large because of the weak perturbative nature of the information extraction, this can be generally overcome by averaging over a vast number of identically prepared states. What makes WM an interesting phenomenon is that by post-selecting the prepared system, the weak value (WV) of an observable may lie well outside of the normal range of eigenvalues of the measurement operator, or even become complex owing to nontrivial interference effect of complex amplitudes. These peculiar features prove to be powerful for deeper understanding of quantum paradoxes and addressing important questions on the foundations of quantum mechanics [2]-[7]. Moreover, the prospect of a WV extending beyond the eigenvalue spectrum, often referred to as amplification [8], has triggered a great deal for metrological applications [9],[10] in the possibility of measuring weak signals by alleviating technical imperfections. Recently this approach has garnered substantial interest and resulted in many astounding observations in birefringence effects [11], electromagnetic pulse propagation [12], optical spin Hall effects [13], transverse beam deflections [14], phase-shift time delays [15], optical angular rotations [16], optical frequency shift [17], and optical nonlinearity at a few photon level [18], to name a few.

On the other hand, precision frequency measurements based upon atomic coherence lie at the heart of many precision measurements such as atomic clocks [19] and optical magnetometry [20]. However, a major challenge is how to attain narrow linewidth but without sacrificing the measurement sensitivity, ascribed by the ratio of the linewidth to the signal-to-noise ratio (SNR). The resonance linewidth is usually limited by the natural lifetime of the quantum states involved or the effective coherence time associated with the atom-light interaction. However, achieving subnatural linewidths does not violate the frequency-time uncertainty relation, as the measurement time can be much longer than the coherence lifetime [21]. Despite several subnatural-linewidth methods [22]-[29] have been put forward in the past, most of them rely on (effectively) filtering out a subgroup of atoms with longer lifetime, and thus inevitably degrade the SNR by a larger factor than that of the linewidth deduction. Even so, as emphasized by Metcalf and Phillips [22], narrow linewidth is still desirable especially when unknown systematic noise deforms the lineshape. Recently, a new resonance method based on measuring intensity correlation in optical fields [30]-[38] has displayed the capability of reducing the correlation-resonance linewidth far below the effective coherence lifetime by 30 times [30]. Unlike other subnatural linewidth spectroscopies [22], this method does not diminish the sensitivity [30] due to the absence of atomic selection, and can also resolve closely-spaced multiple resonance peaks [36]. Unfortunately, these demonstrations so far are all restricted to large resonance linewidths about a few kHz, which has locked its potentials for precision measurements.

In this paper, we introduce the WM approach to the precise measurement of atomic resonance by properly pre- and post-selecting optical states. A subcoherence-lifetime-limited linewidth down to 0.1 Hz has been achieved, together with a magnetic-field sensitivity of 7 fT/$\sqrt{\text{Hz}}$ near DC, for a single laser beam with power as low as 15 μW. This is among the best sensitivity in the ultralow frequency regime particularly near 10 Hz where biomagnetic fields such as that from human brains become relevant [39]. Several unique characteristics appear in our experiment including: prolonged atomic-spin lifetime enabled by the alkene coated vapor cell, working at room temperature, and employing one weak continuous wave (cw) laser for both pumping and probing. Besides these practical advantages, our work further exhibits a few intriguing features beyond the existing WV works, from the theoretical perspective. For instance, in contrast to most of WVs observed to date where light polarization and propagation direction are typically used as the "system" and "pointer", our scheme instead exploits optical frequency and light polarization for the "system" and "pointer". Interestingly, the WV-amplified frequency selection here gives rise to a substantial reduction of the correlation-resonance linewidth. Moreover, the interaction between the system and the pointer is characterized by an imaginary (non-Hermitian) Hamiltonian, resulting in counter observations as compared with conventional Hermitian Hamiltonians [1]-[18]. One practical advantage of this inverse effect is the simplicity to produce an imaginary WV without designing any sophisticated interferometry setup.

The WM protocol is illustrated in a generic three-level Λ-type atomic system (Fig. 1) which is addressed by two circularly-polarized laser fields (left-$\sigma^+$/right-$\sigma^-$) with amplitudes $E_1$ and $E_2$ (developed from one linearly-polarized cw laser) to form electromagnetically induced transparency (EIT). The two-photon detuning Δ is tuned by a magnetic field *B* through shifting the two ground states and correspondingly, the information of *B* is then carried out through polarization and transmission of the output light. The underlying physics of our scheme is similar to that of nonlinear magneto-optical Faraday rotation [40]. By modulating the laser fields at frequency $\omega_m$, the atom-light interaction converts the frequency modulation (FM) to amplitude modulation (AM) for both EIT fields. The conversion for $\sigma^+$- and $\sigma^-$-light is out of phase for a nonzero *B*, and thus results in a small modulation in the polarization state of the light, as schematically shown in Fig.1. After projecting out the atomic evolution (see Supplementary material), for an optically thin medium we arrive at the following effective Hamiltonian for the light (with the dimension of wave vector), which describes the correlation between the optical frequency ("system") and the polarization ("pointer") modeled by the Stokes operator $\sigma_z$:

$$H = -i\xi \sigma_z \otimes (|-\omega_m\rangle\langle 0| + |\omega_m\rangle\langle 0|). \tag{1}$$

Here, $\xi$ is the small perturbative real-value interaction strength which is proportional to the magnetic field and atomic density of the medium, and is also related to modulation parameters. The pointer operator, $\sigma_z = \begin{pmatrix} 1 & 0 \\ 0 & -1 \end{pmatrix}$, assumes the eigenvalue of $+1$ or $-1$ when acting on the corresponding $\sigma^+$-eigenvector $\begin{pmatrix} 1 \\ 0 \end{pmatrix}$ or the $\sigma^-$-eigenvector $\begin{pmatrix} 0 \\ 1 \end{pmatrix}$, respectively. By projecting onto the final state, the expectation value $\langle \sigma_z \rangle$ (as shown in Supplementary material) characterizes the absorption difference between the transmitted $\sigma^+$ and $\sigma^-$ components of the light. In the rotating frame of the modulated laser fields, the FM-AM conversion can be modeled by an operator $|-\omega_m\rangle\langle 0| + |\omega_m\rangle\langle 0|$, with $|\omega_m\rangle\langle\omega_n| = e^{i(\omega_n - \omega_m)t}$. As usual, the frequency vector here obeys the condition $\langle\omega_m|\omega_n\rangle = \delta(\omega_m - \omega_n)$. Note that because the second and higher harmonics are negligible to the first in AM, they are neglected from the interaction Hamiltonian (1). In Eq. (1), $\xi = \frac{\varsigma}{\Gamma}\frac{M}{1+3M^2}\text{Im}[\rho_{cb}]$, with $\varsigma = \frac{n\mu^2}{2\lambda\varepsilon_0\hbar} = \frac{3n\lambda^2}{16\pi^2}\Gamma_0$, $M = \frac{\lambda_m\omega_m}{\Gamma}$ the modulation range ($\lambda_m$ is the modulation depth) normalized to the excited state linewidth $\Gamma$ (phenomenologically taken as the Doppler broadened width), the imaginary part of the ground-state coherence is $\text{Im}[\rho_{cb}] \simeq \frac{(1+M^2)(1+3M^2)\Gamma_p}{[(1+3M^2)\gamma_2 + 2(1+M^2)\Gamma_p]^2}\Delta$, at the limit of small $\Delta (\ll \Gamma_p)$, where $\gamma_2$ is the dephasing rate and $\Gamma_p$ the optical pumping rate for ground state, respectively, and $\Delta$ is the two-photon detuning (equal to the Larmor frequency) which is proportional to $B$; $n$ is the atomic density, $\lambda$ the center wavelength of the input laser, and $\Gamma_0$ is the spontaneous decay rate. Distinct from traditional WM Hamiltonians, the presence of "$i$" in (1) here implies the non-Hermitian nature of the interaction due to the differential absorption of the two EIT fields.

In light of the WM procedure, the atoms are illuminated by an $x$-polarized cw laser, whose initial state is a product of its "pointer" and "system" states: $|\Psi_i\rangle = |\Phi_{pi}\rangle \otimes |\Psi_{si}\rangle = \frac{1}{\sqrt{2}}\begin{pmatrix} 1 \\ 1 \end{pmatrix} \otimes |0\rangle$. After the interaction with the atomic medium of length $L$, governed by the above Hamiltonian, the output final optical state becomes classically entangled between two degrees of freedom in general, $|\Psi_f\rangle \approx (1 - iHL)|\Psi_i\rangle$. By post-selecting the output "system" state $|\Psi_{sf}\rangle$ for the components of central frequency superposed by first side-bands to be measured, however, one could obtain the polarization state of the output light, $|\Phi_{pf}\rangle = \langle\Psi_{sf}|\Psi_f\rangle$, by tracing out the "system". Here, $|\Psi_{sf}\rangle = \frac{1}{\sqrt{(1-D)^2+2}}[(1-D)|0\rangle + |\omega_m\rangle + |-\omega_m\rangle]$, with $D$ a post-selection parameter very close to one, which is nearly orthogonal to $|\Psi_{si}\rangle = |0\rangle$. Alternatively, it can be proved (see Supplementary material) that this post-selection operation is equivalent to subtracting the majority of the DC component of the transmitted light power while retaining the AC components. Note that $|\Psi_{si}\rangle$ and $|\Psi_{sf}\rangle$ take asymmetric forms in contrast to the symmetric ones considered in most of previous experiments [9-18]. After normalization $|\Phi_{pf}\rangle$ can be rewritten as:

$$|\Phi_{pf}\rangle = (1 - \xi L A_W \sigma_z)|\Phi_{pi}\rangle = e^{-\xi L A_W \sigma_z}|\Phi_{pi}\rangle. \qquad (2)$$

Here, the weak value associated with the system observable is defined as

$$A_W = \frac{\langle\Psi_{sf}|(|-\omega_m\rangle\langle 0| + |\omega_m\rangle\langle 0|)|\Psi_{si}\rangle}{\langle\Psi_{sf}|\Psi_{si}\rangle} = \frac{2}{1-D}, \qquad (3)$$

which is a real quantity. It is worth pointing out that in spite of the real $A_W$, the anti-Hermiticity of the interaction Hamiltonian (1) makes the results equivalent to the imaginary WV obtained from a Hermitian one. So far, all realizations of imaginary WVs are indispensable to sophisticated interferometric systems. By constructing a non-Hermitian Hamiltonian, our scheme grants an alternative way to reach the same goal with alleviated complexity of experimental setups. Having $|\Phi_{pf}\rangle$, one can now readily compute $\langle\sigma_z\rangle_{pf} = -2\xi L A_W$, indicating the weak value amplification of the Stokes parameter.

Now, we examine the linewidth $\mathcal{L}$ of our WM-correlation spectroscopy. As studied before, the measurement quantity is the intensity correlation $g^{(2)}(0) = \langle I_1(t)I_2(t)\rangle_T/\sqrt{\langle I_1^2(t)\rangle_T \langle I_2^2(t)\rangle_T}$, between the two EIT fields' output intensities ($I_j(t) = E_j^{(-)}(t)E_j^{(+)}(t)$), at the zero-time lag [30,32]. Here $\langle\cdot\rangle_T$ represents the ensemble average over one modulation period $T = \frac{2\pi}{\omega_m}$. In our WM protocol, for the post-selected $E_1$ and $E_2$ their output intensities are, respectively, replaced by $I_1(t) = I_+(t) - D\langle I_+(t)\rangle_T$ and $I_2(t) = I_-(t) - D\langle I_-(t)\rangle_T$ for the $\sigma^+$ and $\sigma^-$ light. Thanks to the fact that $g^{(2)}(0)$ is bounded between the range of $-1$ and $1$, $\mathcal{L}$ can be deduced by simply seeking the zero-values for the numerator of $g^{(2)}(0)$. Consequently, this is to solve $\langle\left[\frac{I_1(t)-I_2(t)}{\langle I_1(t)\rangle_T+\langle I_2(t)\rangle_T}\right]^2\rangle_T = 1$ (see Supplementary material). Physics becomes now straightforward if one is aware that the argument is essentially the Stokes parameter $\langle\sigma_z\rangle_{pf}$. Based on this observation, the connection between $\mathcal{L}$ and $\langle\sigma_z\rangle_{pf}$ can be rigorously established via $\frac{I_1(t)-I_2(t)}{\langle I_1(t)\rangle_T+\langle I_2(t)\rangle_T} = \frac{I_+(t)-I_-(t)}{(1-D)(I_+(t)+I_-(t))} = 2\langle\sigma_z\rangle_{pf} \times \cos(\omega_m t)$ (see Supplementary material). For standard measurements, $\langle\sigma_z\rangle$ is identical to the Stokes parameter $V$ and thereby suffices the inequality $\langle\sigma_z\rangle \leq 1$. However, the condition of $g^{(2)}(0) = 0$ demands $\langle\sigma_z\rangle = \sqrt{2}$, in opposition to the inequality. It turns out that the WV amplification makes this possible. In such a case, the $g^{(2)}(0)$ linewidth becomes $\mathcal{L} = \frac{\sqrt{2}}{\left|L\left(\frac{\partial\xi}{\partial\Delta}\right)A_W\right|}$, suggesting the feasibility of the anomalous WV-induced narrowing. To assess the performance of the correlation-resonance spectroscopy, one has to look at another critical parameter, the SNR. We can show that under the current WM arrangement, the ultimate SNR, only limited by the photon shot noise, follows the trend of $\frac{2\sqrt{2}}{A_W}\sqrt{n_{\text{ph}}}$ with $n_{\text{ph}}$ being the total photon-number rate (see Method). Because both $\mathcal{L}$ and SNR are inversely proportional to $A_W$, the ultimate sensitivity defined by their ratio has nothing to do with $A_W$, agreeing with the fact that classical experiments cannot breach the quantum limit. Nevertheless, the WM strategy helps to eliminate adverse effect from technical imperfections, and hence permits us to approach the quantum limit for the application to ultra-weak magnetic-field sensing without harsh experimental conditions.

To test our theoretical proposal, in the experiment (Fig. 1) a linearly polarized optical beam was derived from an external cavity diode laser, and then directed into a $^{87}$Rb enriched cylindrical atomic vapor cell (2.5 cm in diameter and 7.5 cm in length) at room temperature (~23°C). With alkene coating [46], the cell was housed inside a four-layer magnetic shield to screen out ambient field. The alkene coating allows atoms to undergo thousands of wall collisions with little demolition of their internal quantum states, giving a zero-power EIT half linewidth of 1 Hz. Inside the shield, a solenoid was used to generate a uniform magnetic field along the laser's propagation direction so that a Zeeman shift $\Delta$ was induced to the two-photon detuning. The input laser, on resonance with the $^{87}$Rb D$_1$ line (795 nm), drives the atomic transitions $|F = 2\rangle \to |F' = 1\rangle$ with its two circular-polarization components, $\sigma^+$ and $\sigma^-$, to form EIT. Their outputs were separately detected by two photo-detectors with gain for analyzing their intensity difference. The laser frequency was modulated at an optimized modulation frequency of 3.03 kHz, and with a modulation range of 250 MHz, by varying the PZT voltage. The residual amplitude modulation (RAM) was reduced by a feedback loop controlling the RF power of an acoustic-optical-modulator in the light stream before the cell. Upon interacting with the atoms, the laser frequency modulation was converted to intensity modulation. Experimentally, we detected the $\omega_m$ components of the AC parts in the $\sigma^+$ and $\sigma^-$ intensities by setting $D$ for the DC background.

We first measure the correlation resonance linewidth $\mathcal{L}$ as a function of the post-selection parameter $D$. For each beam, we extracted the AC components, and the attenuated DC component with an attenuation factor of $(1-D)$ to attain $I_1(t)$ and $I_2(t)$ for computing $g^{(2)}(0)$. Figure 2a gives an example of the $g^{(2)}(0)$ spectrum with a half width at half maximum (HWHM) of 0.1 Hz, ten times smaller than the coherence-lifetime-limited width of 1Hz, obtained by setting $D = 0.9995$. The spectrum, displaying full correlation and anti-correlation, can be well fitted by a Lorentzian profile as predicted in theory. The linewidth can be further reduced but with worse SNR, and the corresponding spectrum also deviated from the Lorentzian lineshape, due to more pronounced adverse effects of technical noises for smaller $(1-D)$ values. The linear dependence of $\mathcal{L}$ on $D$ was well confirmed in Fig. 2b, where $\mathcal{L}$ monotonically decreases along with the gradual increment of $D$. The linewidth $\mathcal{L}$ also depends on the modulation range and laser power, which is optimized in the experiment.

One application of this subnatural linewidth spectroscopy is to sense a weak magnetic field $B$. In order to examine the magnetometer sensitivity, we chose to measure $B$ corresponding to the zero value of $g^{(2)}(0)$ where the $g^{(2)}(0)$ vs. $B$ curve has a large slope. The $g^{(2)}(0)$ value is obtained in one modulation period $T$. The procedure was repeated and continuously measured for one second to produce $1/T$ copies of the correlation value, and then a Fourier transform was done to produce the noise spectra of $g^{(2)}(0)$. In order to obtain the sensitivity, we then measured the slope of the $g^{(2)}(0)$ curve at the measured $B$, but with an alternating magnetic field at various frequencies (see Fig. 3a). The slope was used to convert the standard deviation of $g^{(2)}(0)$ into the deviation of $B$, rendering the sensitivity in unit of $\text{fT}/\sqrt{\text{Hz}}$. It turns out that our magnetometer is highly sensitive to low-frequency magnetic fields in the range of few Hz to tens of Hz. As an example, the measured frequency response well fitted to the function $BW/\sqrt{BW^2 + f^2}$ is presented in Fig. 3b by looking at the slope of $g^{(2)}(0)$ resonance at HWHM for the applied $B$ at alternating frequencies. Here $BW$ stands for a fitting parameter associated with the linewidth. Due to the small fluctuation of the residual ambient magnetic field as well as of the applied $B$, we found experimentally that the optimal single-shot measurement time is one modulation period. Although $\mathcal{L}$ decreases with larger $D$, a larger $D$ could adversely aggravate the effect of technical noise on the $g^{(2)}(0)$ value and results in the reduction of the SNR. Therefore, an optimized value for $D$ is anticipated in practice. Our demonstration approved such an expectation. As shown in Fig. 4, the optimal sensitivity of 7 $\text{fT}/\sqrt{\text{Hz}}$ (in the 10 ~ 20 Hz range) ends up around $D = 0.995$, corresponding to a half width of about 1 Hz for $g^{(2)}(0)$. The bandwidth of the magnetometer is relatively small, mainly determined by the optical pumping rate of the system, which is 13 Hz for laser power of 15 μW. The sensitivity in the 5 ~ 10 Hz range is still below 10 $\text{fT}/\sqrt{\text{Hz}}$, which is highly applicable to low-frequency biomagnetic field sensing, and comparable to the best in this frequency range so far to our knowledge for a single-channel magnetometer operating at 200°C [42]. The room temperature and low laser-power operation condition further makes this scheme attractive for situations demanding low power consumption [43].

It is known that the WM approach can reduce the negative effects of certain technical noise on the measurement, mainly by carefully engineering the overlap between the initial and final "system" states. Since a typical noise source in most FM experiments is the RAM in the laser, we have designed a feedback loop to reduce the RAM level by 25 dB. To investigate the effect of RAM in a controlled way, we added common-mode random white intensity noise at the two outputs of varied amount. Similar to previous WM research, the sensitivity becomes worse when $|\Psi_{si}\rangle$ and $|\Psi_{sf}\rangle$ are orthogonal, i.e., when $D = 1$ and the DC part of the output intensity is zero. In such a case, the deleterious correlated noise including RAM and RIN (random intensity noise) from the laser becomes prominent and uncontrollable. These noise cause random fluctuations in the correlated $g^{(2)}(0)$ signal. As a matter of fact, such common-mode noise widens the correlation resonance linewidth $\mathcal{L}$ and lowers the sensitivity considerably. As shown in Fig. 4, we measured the magnetometer sensitivity for different values of $D$, at different amounts of added RAM. There is a pronounced sensitivity degradation as $D$ approaches one, consistent with the above analysis. When $D$ is lessened, however, the sensitivity becomes less sensitive to RAM and similar to the case without added RAM, because the projected DC component dominates over the noise amplitude. For small $D$, the offset of the experimental data from theoretical simulations is a result of the failure of the approximation that the AC-component amplitudes are linearly proportional to $\Delta$. Interestingly, even when the input linear polarization is not pure, this WM method is more immune to the RAM noise than normal intensity difference measurement, as shown in Fig.5.

We further investigated the shot-noise-limited magnetometer sensitivity, which shall be independent of $D$ as mentioned above. The best sensitivity of 7 fT/$\sqrt{\text{Hz}}$ was achieved in the experiment, close to the calculated 3 fT/$\sqrt{\text{Hz}}$ taking into account the frequency response near 10 Hz. A careful noise analysis has shown that the gap between the present sensitivity and the photon-shot-noise-limit is mainly from the magnetic field noise from both the environment and the applied field itself. Despite the former could be eliminated in a gradiometer by multiple-channel operation [44], this is not applicable to a coated cell due to motional averaging of the magnetic field across the cell volume. We have also calculated the photon-shot-noise limited sensitivity for NMOR and EIT using similar parameter settings and found them to be similar to this method. The estimated atomic shot noise limit in this system is 0.2 fT/$\sqrt{\text{Hz}}$ at DC.

The approach of post-selection in the frequency domain is closely related to the well-known phase-sensitive lock-in detection technique [42], widely applied in measurements to extract small signals from the background noise, by the modulation-demodulation procedures. There, using the WM language, the input and output signals are completely orthogonal, because the input laser has no intensity modulation, the output has it through FM-AM or AM-AM conversion, while the DC component is completely discarded in the demodulation process. Here, in contrast, the key is to keep a controllable amount of DC signal in the output to maintain the small non-orthogonality between the input and output, which is essential for all WM-post-selection experiments. The smaller the overlap between $|\Psi_{si}\rangle$ and $|\Psi_{sf}\rangle$, the larger the signal and visibility are. In our scheme, although the anti-correlation signal originates from a nonzero magnetic field, $g^{(2)}(0)$ would always be nearly $-1$ without the retained DC component. Alternatively, the DC component here provides a comparison basis in the measurement and differentiates this method from traditional lock-in detection.

Another interesting aspect of our scheme is to avoid backaction. Originated from the photon shot noise, quantum noise randomizes the phase of the ground-state atomic-spin through off-resonant atom-light coupling (AC-stark shift), and it practically exists nearly everywhere including in our scheme. As pointed out in [46], such backaction leads to squeezed light but degrades the measurement of polarization rotation in the equator plane of the Poincaré sphere. For instance, the broadened Stokes component (nearly the anti-squeezed component of light) can be as large as 10 dB [47], indicating a sensitivity 3.3 times above the shot noise limit. However, our method measures along the longitudinal direction of the Poincaré sphere, which is free from such backaction. Though our current experiment does not suffer from the backaction much due to the low optical depth at room temperature, this backaction problem will become apparent at higher atomic density and shall be carefully considered.

We have used WV amplification to narrow the resonance linewidth in warm $^{87}$Rb vapor down to the level absent from previous capability. The scheme exhibits a number of intriguing features that distinguish our work from the existing research on WM. These include the selections on "system" and "pointer", the anti-Hermiticity of the interaction Hamiltonian, and ingenious relation between the Stokes parameter and the $g^{(2)}(0)$ measurement. From a theoretical perspective, our results open a door for disseminating WM to non-Hermitian physics where nontrivial phenomena are expected to be discovered. Notwithstanding the classical light, the simplicity and good-performance of the scheme makes our magnetometer very promising for real applications. The low power consumption is also beneficial for avoiding cell heating and multiple lasers, thanks to the resonant interaction further enhanced by the long-lived coherence and the WM approach. Despite the drawback of the WM-induced resonance narrowing approach for magnetometry is to sacrifice the dynamical range for detectable magnetic field, this could be overcome by applying an offset field. Nevertheless, a line narrowing technique is still important for spectroscopy, as emphasized in [22]. Future development will incorporate quantum enhancement by including the squeezed light or the squeezed atomic spin.

**Acknowledgements**

We are grateful to Irina Novikova, Eugeniy Mikhailov, and Kaifeng Zhao for helpful discussions.

**List of Figures**

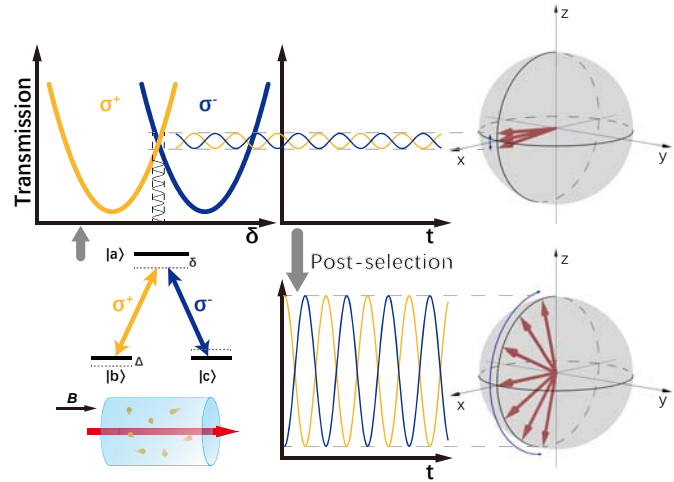

**Figure 1 | Principle of WM-enhanced correlation spectroscopy for weak, low-frequency magnetic-field sensing in a $^{87}$Rb vapor cell at room temperature.** Schematic of correlation spectroscopy assisted by pre- and post-selecting frequency components of the two EIT fields, $\sigma^+$ and $\sigma^-$, in a three-level Λ configuration. Such pre- and post-selection essentially makes an anomalous amplification on the expectation value of the Stokes polarization operator, $\langle \sigma_z \rangle_{pf}$. Here, δ is the averaged one-photon detuning.

Δ is the two-photon detuning which is generated by shifting the energy levels with a magnetic field *B* via Zeeman shift.

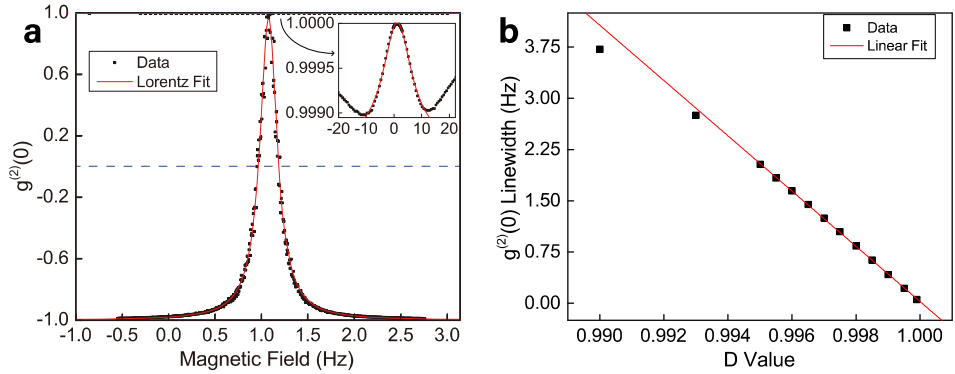

**Figure 2 | Representative correlation-resonance spectrum of $g^{(2)}(0)$ and its linewidth $\mathcal{L}$. a,** Exemplar of the $g^{(2)}(0)$ spectrum with the HWHM of about 0.1 Hz, which is $1/10$ of the coherence-lifetime-limited half width of 1 Hz, by setting *D* = 0.9995. The inset shows the spectrum without post-selection, i.e., *D* = 0. The zero applied field is not at the spectra center due to residual stray offset field inside the shield. **b,** linear $\mathcal{L}$ dependence (FWHM) on the projection parameter *D* for an input *x*-polarized cw laser of power 11 μW, in good agreement with the trend predicted by theory.

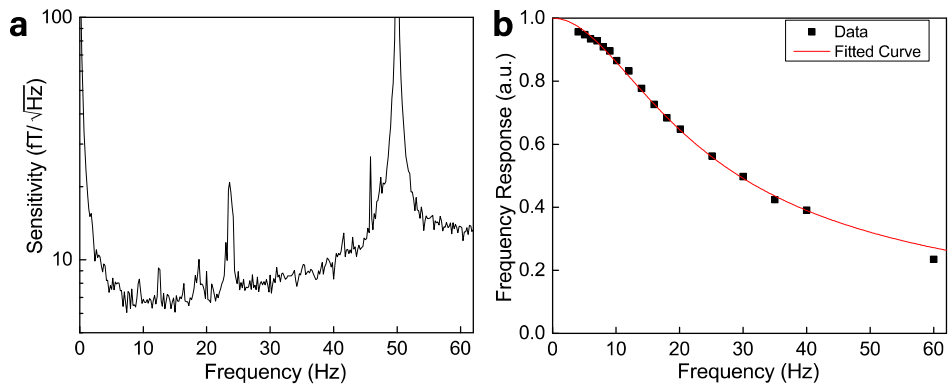

**Figure 3 | Sensitivity and bandwidth of the magnetometer for sensing low-frequency magnetic fields sensing. a,** The sensitivity derived from the $g^{(2)}(0)$ noise spectrum as well as the measured frequency response. Here, the $g^{(2)}(0)$ noise spectrum was obtained from Fourier transforming 3030 copies of $g^{(2)}(0)$ values at HWHM (i.e. $g^{(2)}(0) \sim 0$) which were continuously measured within one second with each point taken in one modulation period. **b,** By normalizing with respect to DC background, the measured frequency response (or the slope of the $g^{(2)}(0)$ resonance at HWHM) for the applied AC magnetic field at varying frequencies. The fitting curve follows the function $BW/\sqrt{BW^2 + f^2}$ with $BW = 16$ Hz. Experimental parameters here: laser power of 15 μW and *D* = 0.995.

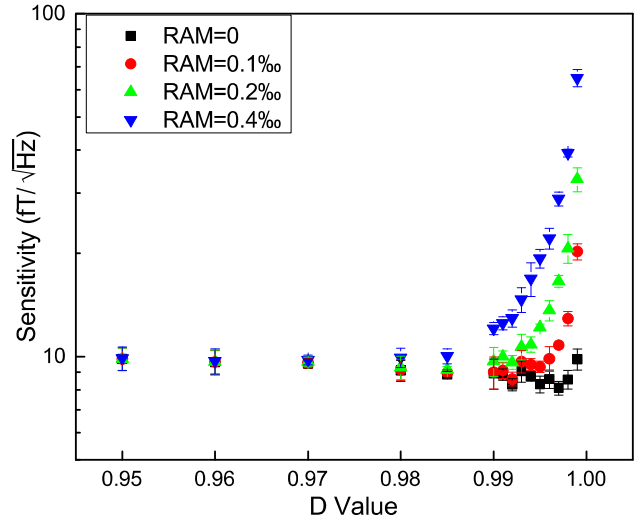

**Figure 4 | Dependence of sensitivity on the projection parameter *D* investigated by changing the added amounts of residual amplitude modulation (RAM).** In the experiment, different amounts of common mode intensity noise were added to the output intensities to mimic RAM. The magnetic field sensitivity was the averaged sensitivity between 12 Hz and 22 Hz by choosing a proper *D* value to optimize the sensitivity. The sensitivity was then measured when a magnetic field corresponds to the largest slope of the $g^{(2)}(0)$ resonance. Experimental parameters are same as those in Fig. 3 except for varying *D* values.

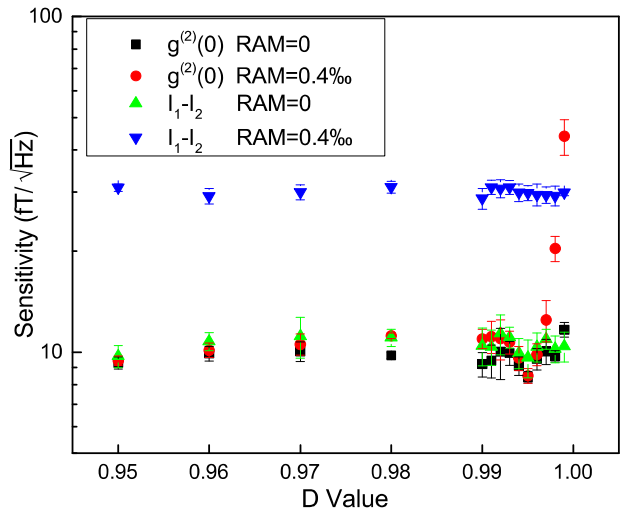

**Figure 5 | Comparison of the sensitivities achieved via the $g^{(2)}(0)$ method and the normal intensity-difference measurement with or without added RAM for an elliptically-polarized input laser beam (of power 15 μW and ellipticity 1:2).** With added RAM, and for D less than 0.996, the intensity difference method is inferior to the correlation method. Without RAM, the two methods are comparable.